\documentclass[9pt,twocolumn,twoside]{opticajnl}
\newcommand{\publishedDOI}{}   

\journal{opticajournal}          
\setboolean{shortarticle}{true}  
\setboolean{displaycopyright}{true}
\renewcommand{\copyrightstatement}{\par\smallskip\noindent
  \textcopyright\ 2026 Optica Publishing Group. One print or electronic copy
  may be made for personal use only. Systematic reproduction and distribution,
  duplication of any material in this paper for a fee or for commercial
  purposes, or modifications of the content of this paper are prohibited.
  \ifdefempty{\publishedDOI}{%
    Accepted for publication in Optics Letters; the published version will be
    available from Optica Publishing Group.}{%
    Published version: \url{https://doi.org/\publishedDOI}.}}
\doi{}
\dates{Accepted manuscript, Optics Letters, November 2026. arXiv version 1.}

\usepackage{lineno}

\title{A universal modal-noise limit for multimode photonic systems}

\author{K. Madhav}

\affil{Astrophotonics (innoFSPEC), Leibniz-Institut f\"ur Astrophysik Potsdam (AIP),
An der Sternwarte 16, 14482 Potsdam, Germany}

\affil{\textsuperscript{*}kmadhav@aip.de}

\begin{abstract}
Modal noise limits the photometric and spectroscopic stability of every
multimode-to-single-mode photonic system, yet is quantified device by device.
For isotropic modal excitation at fixed input power, the fractional power
fluctuation of any linear multimode system is shown to be fixed by one number:
the participation ratio $M_{\mathrm{eff}}$ of the eigenvalues of
$K=T^{\dagger}WT$. Monte Carlo for spatially filtered fibres, lossy lanterns
and integrated mode converters confirms the law, which recovers the classical
speckle limits. Extending $K$ to $K(\lambda)$ gives the spectral correlation of
modal noise and a closed-form velocity error, maximal when the correlation
width equals the line width.
\end{abstract}

\begin{document}

\maketitle

\section{Introduction}
Any optical system that collects light into a multimode waveguide and then
filters it, spatially or modally, converts a fluctuation of the modal
excitation into a fluctuation of the detected power. This is modal
noise~\cite{epworth1979,rawson1980}. Because the modal excitation responds to seeing, guiding, source motion,
temperature and mechanical stress, modal noise is multiplicative, does not
average down with integration time, and does not appear in the photon-noise
budget.

Its consequences are felt in fields that rarely cite one another. In fibre-fed
astronomical spectrographs it sets the floor on radial-velocity precision and
line-profile
stability~\cite{baudrand2001,lemke2011,mahadevan2014,halverson2015b,oliva2019},
which is why scrambling optics, non-circular fibres and double scramblers are
standard~\cite{spronck2015,bouchy2013,halverson2015}. In
space-division-multiplexed telecommunications the same physics appears as
mode-dependent loss (MDL) and caps MIMO channel
capacity~\cite{ho2011,winzer2011}. In multimode-fibre spectrometers the speckle
pattern is the signal rather than the
noise~\cite{redding2013,redding2014,cao2017}. Photonic lanterns, reformatters
and integrated mode converters sit at the centre of all of
these~\cite{leonsaval2005,noordegraaf2009,noordegraaf2010,leonsaval2014,birks2015,blandhawthorn2017,norris2020,pike2020}.

What is missing is a device-independent statement of how much modal noise a
given system must have. Existing photonic treatments model a specific fibre,
scrambler or lantern, and their results do not transfer. A closely related
quantity is known in two other settings: the effective number of incoherently
summed speckle patterns~\cite{rawson1980}, and the participation number of the
eigenvalues of $tt^{\dagger}$ governing total-transmission variance in random
media~\cite{davy2013}. This quantity, evaluated on the transmittance
eigenvalues of the device itself and exact at finite mode count, is shown here
to be the governing scalar for engineered multimode photonic systems, and to
control the wavelength correlation of the noise and its conversion into a
velocity error as well.

\section{Transfer-matrix formulation}
Let $a\in\mathbb{C}^{M}$ hold the complex amplitudes of the $M$ modes excited
at the input and $b\in\mathbb{C}^{N}$ the output amplitudes, related by the
transfer matrix, $b(\lambda)=T(\lambda)a(\lambda)$. The measurement is a
quadratic form in $b$. Writing the mode-dependent detection and loss of
everything downstream, a spatial filter, a slit, a detector footprint or a
subset of lantern ports, as a Hermitian positive-semidefinite operator
$W(\lambda)$, the detected power $P=b^{\dagger}Wb$ becomes
\begin{equation}
P=a^{\dagger}Ka,
\qquad
K(\lambda)\equiv T^{\dagger}(\lambda)\,W(\lambda)\,T(\lambda).
\label{eq:K}
\end{equation}
Every property of the system that can affect modal noise is contained in the
single Hermitian operator $K$, whose eigenvalues
$\kappa_{1}\ge\dots\ge\kappa_{M}\ge0$ are the mode-dependent transmittances of
the complete multimode-to-detector chain. Nothing has been assumed about the
physical realisation of $T$: it may describe a multimode, few-mode or multicore
fibre, a photonic lantern, a mode multiplexer, a reformatter, an integrated
spectrograph, or a cascade of these.

For the input statistics the incident power $P_{0}$ is taken fixed and the
modal excitation isotropic,
\begin{equation}
a=\sqrt{P_{0}}\,u,\qquad u^{\dagger}u=1,
\label{eq:iso}
\end{equation}
with $u$ uniform on the unit sphere of $\mathbb{C}^{M}$. This is the
maximum-entropy state consistent with a fully filled, strongly mixed guide and
a stable source, and it isolates modal noise from photon noise and from
throughput drift by construction.

\section{The universal law}
With $\langle u_{i}u_{j}^{*}\rangle=\delta_{ij}/M$ and
$\langle u_{i}u_{j}^{*}u_{k}u_{l}^{*}\rangle=
(\delta_{ij}\delta_{kl}+\delta_{il}\delta_{kj})/[M(M+1)]$,
\begin{equation}
\langle P\rangle=\frac{P_{0}}{M}\,\mathrm{Tr}\,K,
\quad
\mathrm{Var}(P)=\frac{P_{0}^{2}\!\left[M\,\mathrm{Tr}(K^{2})-(\mathrm{Tr}K)^{2}\right]}
{M^{2}(M+1)}.
\label{eq:moments}
\end{equation}
Introducing the participation ratio of the mode-dependent transmittances,
$M_{\mathrm{eff}}\equiv(\mathrm{Tr}K)^{2}/\mathrm{Tr}(K^{2})=
(\sum_{m}\kappa_{m})^{2}/\sum_{m}\kappa_{m}^{2}$,
the two moments collapse into a single relation,
\begin{equation}
\boxed{\;
\frac{\sigma_{P}}{\langle P\rangle}
=\left[\frac{M/M_{\mathrm{eff}}-1}{M+1}\right]^{1/2}.
\;}
\label{eq:law}
\end{equation}
This is the central result: the modal-noise floor of an arbitrary linear
multimode system depends on the system only through $M_{\mathrm{eff}}$, the
number of modes that effectively participate in transmission. It is the
participation ratio used to count states in disordered
systems~\cite{thouless1974}, and the form recurs across optics: it is the effective number of incoherently summed speckle
patterns of unequal mean~\cite{rawson1980}, and the participation number of the
eigenvalues of $tt^{\dagger}$ that fixes total-transmission variance in random
media~\cite{davy2013}. Equation~(\ref{eq:law}) is its statement for the
transmittance eigenvalues of a designed photonic system, exact at finite $M$
under fixed input power; for $M\gg1$,
$(\sigma_{P}/\langle P\rangle)^{2}\to M_{\mathrm{eff}}^{-1}$. Keeping
$W$ explicit rather than absorbing $W^{1/2}$ into $T$ is a matter of use, not
generality: $T$ is measured once per device, whereas $W$ changes with every
spectrograph, detector or port selection behind it.

Three limits fix the interpretation. For a mode-independent system, $K=\eta I$,
$M_{\mathrm{eff}}=M$ and $\sigma_{P}=0$ exactly: random modal redistribution
cannot change the output power, however many modes are present. For a rank-$r$
projector, an ideal filter passing $r$ of $M$ modes, $M_{\mathrm{eff}}=r$ and,
for $r\ll M$, $\sigma_{P}/\langle P\rangle\simeq r^{-1/2}$, the classical
result for speckle with $r$ degrees of
freedom~\cite{goodman1981,goodman2020}. For $r=1$, single-mode filtering of a
highly multimode field, Eq.~(\ref{eq:law}) gives $[(M-1)/(M+1)]^{1/2}\to1$, the
unit contrast of fully developed speckle. The theory reproduces the known
limits and interpolates between them continuously. It is also bounded:
$\sigma_{P}/\langle P\rangle<1$ for any system whatsoever. Since
$M_{\mathrm{eff}}$ depends on $K$ only through its spectrum, and the isotropic
ensemble is itself invariant under unitary transformations, a relabelling of
modes or a lossless unitary acting on the input, $T\to TU$, leaves the floor
unchanged. A unitary inserted between the device and a fixed mode-selective
detector, $T\to UT$, is not of this kind: it replaces $W$ by $U^{\dagger}WU$
and can alter the spectrum of $K$, which is how a scrambler placed before a
slit acts. Mode scrambling helps only to the extent that it flattens the
transmittance spectrum.

The isotropic assumption can be relaxed where it matters most. If the launch
excites only an $m$-dimensional subspace $\mathcal{S}$, the case of an
under-filled injection or a mode-selective launch, then
$P=P_{0}u^{\dagger}K_{\mathcal{S}}u$ with
$K_{\mathcal{S}}=P_{\mathcal{S}}^{\dagger}KP_{\mathcal{S}}$ the compression of
$K$ onto $\mathcal{S}$, and Eq.~(\ref{eq:law}) holds with $M\to m$ and
$K\to K_{\mathcal{S}}$. Under-filling therefore raises the floor even though the
device is unchanged, which is the quantitative content of the long-standing
requirement that fibre-fed instruments fill the input numerical aperture.

\section{Numerical verification}
\begin{figure*}[t]
\centering
\includegraphics[width=\textwidth]{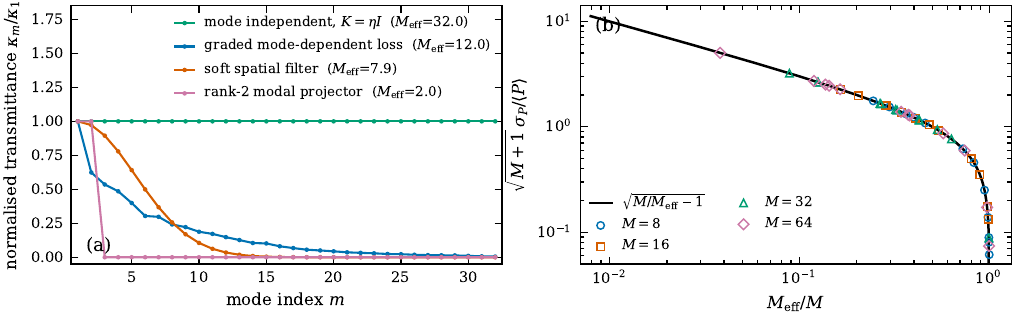}
\caption{Universality of the modal-noise floor. (a) Normalised mode-dependent
transmittance spectra $\kappa_{m}$ of four systems with $M=32$ but very
different participation ratios. (b) Monte-Carlo modal noise, scaled by
$\sqrt{M+1}$, against $M_{\mathrm{eff}}/M$ for $M=8$, $16$, $32$ and $64$ and
for randomly generated spatial filters, lossy lanterns and mode converters. All
points collapse onto the parameter-free prediction of Eq.~(\ref{eq:law}), solid
line.}
\label{fig:fig1}
\end{figure*}

Figure~\ref{fig:fig1} tests Eq.~(\ref{eq:law}) directly: modal states are drawn
from the ensemble of Eq.~(\ref{eq:iso}) and $P=P_{0}u^{\dagger}Ku$ is evaluated
without approximation. Panel (a) shows how different the underlying
transmittance spectra can be; panel (b) plots
$\sqrt{M+1}\,\sigma_{P}/\langle P\rangle$ against $M_{\mathrm{eff}}/M$, which
removes the explicit $M$ dependence of Eq.~(\ref{eq:law}). Systems with $M=8$
to $64$, spanning two decades in $M_{\mathrm{eff}}/M$, fall on the single
master curve with no fitted parameter.

\begin{figure}[t]
\centering
\includegraphics[width=\columnwidth]{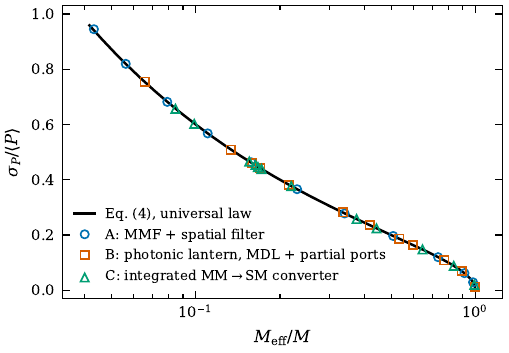}
\caption{Three unrelated device classes, one law. Monte-Carlo modal noise for
(A) a multimode fibre with a soft spatial filter, (B) a photonic lantern with
mode-dependent loss and partial port use, and (C) an integrated
multimode-to-single-mode converter with strongly coupled MDL, all with $M=24$.
Solid line: Eq.~(\ref{eq:law}).}
\label{fig:fig2}
\end{figure}

Figure~\ref{fig:fig2} makes the same point across device classes. Three
unrelated systems are swept over their own design parameters: (A) a strongly
mixing multimode fibre with a soft spatial filter, sweeping the filter width;
(B) a photonic lantern with mode-dependent loss~\cite{noordegraaf2009,noordegraaf2010}
in which only a subset of the single-mode ports is relayed, sweeping loss and
port count; and (C) an
integrated mode converter whose MDL follows the strong-coupling model
$T=U_{2}\exp(H/2)U_{1}$ with $H$ traceless Gaussian Hermitian~\cite{ho2011},
sweeping the MDL spread. The three families have entirely different $K$ spectra
and physical origins, and lie on one curve.

\section{Spectral modal noise}
Because $K$ inherits the wavelength dependence of $T$ and $W$, the same
formalism gives the spectral structure of the noise. For one modal state
observed at two wavelengths, Eq.~(\ref{eq:moments}) generalises to
\begin{equation}
\mathrm{Cov}(P_{1},P_{2})=
\frac{P_{0}^{2}\!\left[M\,\mathrm{Tr}(K_{1}K_{2})-\mathrm{Tr}K_{1}\mathrm{Tr}K_{2}\right]}
{M^{2}(M+1)},
\label{eq:cov}
\end{equation}
with $K_{i}=K(\lambda_{i})$. Consider the generic architecture in which a
multimode section of length $L$ is followed by a wavelength-independent mixing
and detection stage, $T(\lambda)=C\Phi(\lambda)$ with
$\Phi=\mathrm{diag}[\exp(\mathrm{i}\beta_{m}(\lambda)L)]$, so that
$K(\lambda)=\Phi^{\dagger}G\Phi$ with $G=C^{\dagger}WC$ fixed. Then
$\mathrm{Tr}\,K$ is wavelength independent, the mean throughput does not
disperse, and
\begin{equation}
\mathrm{Tr}\!\left[K(\omega_{1})K(\omega_{2})\right]
=\sum_{m,n}\left|G_{mn}\right|^{2}\cos\!\left(\tau_{mn}\Delta\omega\right),
\label{eq:structure}
\end{equation}
where $\tau_{mn}$ is the differential group delay between modes $m$ and $n$.
The spectral correlation function of modal noise is therefore the Fourier
transform of the differential mode-delay distribution weighted by
$|G_{mn}|^{2}$. The equal-weight version of this relation, and the resulting
decorrelation bandwidth, were obtained for a planar multimode guide by Rawson
\textit{et al.}~\cite{rawson1980} and exploited for bandwidth metrology by
Moslehi \textit{et al.}~\cite{moslehi1983}. Equation~(\ref{eq:structure})
supplies the weights $|G_{mn}|^{2}$, which were previously unspecified, and
holds for an arbitrary mixing and detection stage.

\begin{figure*}[t]
\centering
\includegraphics[width=\textwidth]{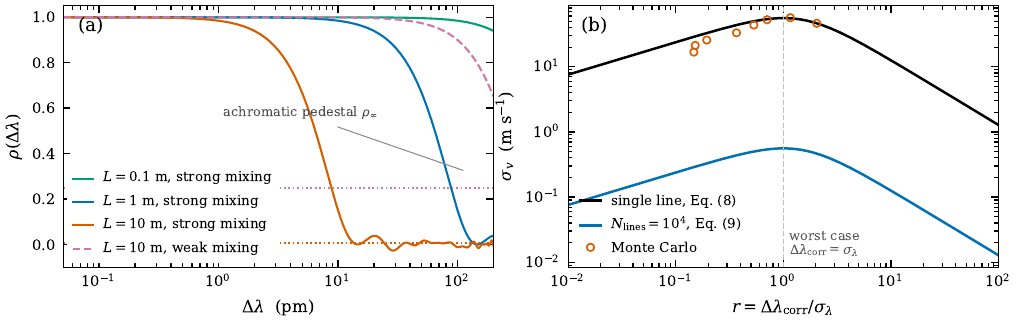}
\caption{Spectral modal noise and its conversion to velocity error.
(a) Normalised spectral correlation function from Eq.~(\ref{eq:structure}) for
a step-index multimode fibre of length $L$ at $\lambda_{0}=1550$~nm, in the
strong- and weak-mixing regimes; dotted lines mark the achromatic pedestals of
Eq.~(\ref{eq:pedestal}). (b) Velocity error against
$r=\Delta\lambda_{\mathrm{corr}}/\sigma_{\lambda}$ for
$\sigma_{\varepsilon}=0.1$ and $R=10^{5}$, for one line and for $10^{4}$
averaged lines. Circles: Monte Carlo using the exact non-Gaussian correlation
function, which accounts for the small-$r$ offset.}
\label{fig:fig3}
\end{figure*}

Equation~(\ref{eq:structure}) also exposes a feature a phenomenological
treatment misses. The $m=n$ terms carry no modal phase, so they never
decorrelate: the normalised correlation function $\rho(\Delta\lambda)$ tends
not to zero but to an achromatic pedestal
\begin{equation}
\rho_{\infty}=\frac{M\sum_{m}G_{mm}^{2}-(\mathrm{Tr}G)^{2}}
{M\sum_{m,n}|G_{mn}|^{2}-(\mathrm{Tr}G)^{2}},
\label{eq:pedestal}
\end{equation}
the fraction of modal noise that is a grey, wavelength-independent throughput
fluctuation. Incomplete decorrelation of this kind was reported in early
measurements~\cite{rawson1980}; Eq.~(\ref{eq:pedestal}) identifies its origin
and fixes its value. Modal noise thus splits into a grey component, harmless to
velocity measurement but damaging to spectrophotometry, and a chromatic
component with the opposite behaviour. Strong mixing drives
$\rho_{\infty}\to0$ and moves all the noise into the chromatic channel; weak
mixing does the reverse. Figure~\ref{fig:fig3}(a) shows both regimes.

\section{Transfer to velocity error}
Write the measured spectrum as $I(\lambda)=S(\lambda)[1+\varepsilon(\lambda)]$,
with $\sigma_{\varepsilon}$ the fractional modal noise per spectral element
from Eq.~(\ref{eq:law}) and correlation function $\rho$ from
Eq.~(\ref{eq:structure}). The intensity-weighted centroid is displaced by
$\delta\lambda_{c}=\int(\lambda-\lambda_{0})S\varepsilon\,\mathrm{d}\lambda/
\int S\,\mathrm{d}\lambda$. For a Gaussian line of rms width
$\sigma_{\lambda}$ and a Gaussian $\rho$ of rms width
$\Delta\lambda_{\mathrm{corr}}$ this integrates in closed form to
\begin{equation}
\mathrm{Var}(\delta\lambda_{c})
=\sigma_{\varepsilon}^{2}\sigma_{\lambda}^{2}\,
\frac{r}{\left(r^{2}+2\right)^{3/2}},
\qquad r\equiv\frac{\Delta\lambda_{\mathrm{corr}}}{\sigma_{\lambda}},
\label{eq:centroid}
\end{equation}
with velocity error
$\sigma_{v}=(c/\lambda_{0})\mathrm{Var}(\delta\lambda_{c})^{1/2}$.

Equation~(\ref{eq:centroid}) is a design rule. It vanishes as $r^{1/2}$ for
$r\ll1$, because rapidly decorrelating modal noise averages within the line,
and as $r^{-1}$ for $r\gg1$, because spectrally flat modal noise scales the
line without moving it. It is maximal at $r=1$, where
$\sigma_{v}^{\max}=3^{-3/4}(c/\lambda_{0})\sigma_{\varepsilon}\sigma_{\lambda}$.
The worst case is not the noisiest system but the one whose modal-noise
correlation width matches its resolution element, and that is where fibre-fed
high-resolution spectrographs operate: a $10$~m step-index fibre with
$\mathrm{NA}=0.22$ has $\Delta\lambda_{\mathrm{corr}}\approx2$~pm at
$1550$~nm, while $R=10^{5}$ gives $\sigma_{\lambda}\approx6.6$~pm, so
$r\approx0.3$, which already returns $72\%$ of the worst-case error. Real
instruments sit near the peak of Eq.~(\ref{eq:centroid}), not in either benign
limit. Averaging $N_{\mathrm{lines}}$ lines that see independent modal
noise, which holds when the line spacing exceeds
$\Delta\lambda_{\mathrm{corr}}$, reduces the error by
$N_{\mathrm{lines}}^{-1/2}$,
\begin{equation}
\sigma_{v}\simeq\frac{c}{\lambda_{0}}
\frac{\sigma_{\varepsilon}\sigma_{\lambda}}{\sqrt{N_{\mathrm{lines}}}}
\left[\frac{r}{(r^{2}+2)^{3/2}}\right]^{1/2}.
\label{eq:nlines}
\end{equation}
For $\sigma_{\varepsilon}=0.1$, $R=10^{5}$ and $N_{\mathrm{lines}}=10^{4}$ this
gives $\sigma_{v}\sim0.5$~m\,s$^{-1}$ near the worst case, consistent with
modal-noise floors reported for fibre-fed
instruments~\cite{baudrand2001,mahadevan2014,halverson2015b}.

\section{Discussion}
The practical content of Eq.~(\ref{eq:law}) is that modal-noise engineering is
eigenvalue engineering. Measuring the transfer matrix of a
device~\cite{popoff2010,rotter2017} and forming $K=T^{\dagger}WT$ with the
actual downstream $W$ yields the achievable stability before an instrument is
built. A mode-selective lantern~\cite{leonsaval2014} with uniform port
transmission is preferable to a lower-loss device with a spread of port
transmissions, because $M_{\mathrm{eff}}$, not throughput, sets the noise;
equally, discarding lantern ports or truncating a reformatted pseudo-slit costs
stability faster than flux, since $\sigma_{P}/\langle P\rangle$ scales as
$M_{\mathrm{eff}}^{-1/2}$ while throughput scales as $M_{\mathrm{eff}}$. The
same quantity is the MDL figure of merit in space-division-multiplexed
links~\cite{ho2011,winzer2011}, so a lantern characterised for a spectrograph
and one characterised for a mode multiplexer are described by one number. For speckle spectrometers the sign reverses: those instruments want small
$M_{\mathrm{eff}}$ and short $\Delta\lambda_{\mathrm{corr}}$, and
Eqs.~(\ref{eq:law}) and (\ref{eq:structure}) give that trade
directly~\cite{redding2013,cao2017}.

Equation~(\ref{eq:moments}) also bounds the adaptive conversion of multimode
light into a single mode. For an input described by a modal density matrix
$\rho$, $\mathrm{Tr}\,\rho=1$, the mean detected power is
$\langle P\rangle=P_{0}\mathrm{Tr}(K\rho)$, of which the isotropic average
in Eq.~(\ref{eq:moments}) is the case $\rho=I/M$. For a single-mode output
$K=\lvert t\rangle\langle t\rvert$ with $\lVert t\rVert\le1$ for any passive
device, so $\langle P\rangle\le P_{0}\lambda_{\max}(\rho)$ whatever $t$ is
chosen to be. A spatially coherent input has $\lambda_{\max}=1$ and can be
steered into one mode by a reconfigurable $T$ with of order $2(m-1)$ controls,
which is wavefront shaping~\cite{vellekoop2007,popoff2010}; a spatially
incoherent input filling $m$ modes has $\lambda_{\max}=1/m$, which no passive
or adaptive linear device exceeds. A single setting moreover holds only within
the correlation width of Eq.~(\ref{eq:structure}). In-waveguide adaptive
converters driven by a laser through a multimode fibre~\cite{deng2026} belong
to the first case, and their reported gains are of the order expected from the
number of controls; what cannot be inferred from such a demonstration is the
performance for an extended or broadband source, which the second case bounds
at $1/m$ per setting.

Two limitations should be stated. The isotropic ensemble is an upper bound on
modal disorder; a partially stabilised system lies below Eq.~(\ref{eq:law}),
and the subspace compression above gives the correction when the launch is
restricted rather than stabilised. Second, this is a single-realisation floor: if the modal state decorrelates
$N_{\mathrm{t}}$ times within an exposure, from seeing, agitation or deliberate
modulation, the measured noise falls as $N_{\mathrm{t}}^{-1/2}$ while
$M_{\mathrm{eff}}$ is unchanged. Mechanical agitation buys averaging, not a
better device.

In summary, one scalar computed from the transfer matrix and the downstream
detection operator fixes the modal-noise floor of a multimode photonic system,
the wavelength scale over which that noise persists, and the velocity error it
produces. The result is device independent, and applies to fibres, lanterns,
reformatters and integrated multimode systems alike.

\begin{backmatter}
\bmsection{Funding} Content in the funding section will be generated entirely
from details submitted to Prism.

\bmsection{Disclosures} The author declares no conflicts of interest.

\bmsection{Data availability} The code that generates every result and figure
in this Letter is included with the manuscript source files and is available
from the author on request.
\end{backmatter}

\bibliography{references}

\bibliographyfullrefs{references}

\end{document}